\documentclass[twocolumn]{aastex631}

\newcommand{\caHK}{\mbox{Ca\,{\textsc{ii}}~H\,\&\,K\,}}

\begin{document}

\title{Detectability of Solar Rotation Period Across Various Wavelengths}

%% The new \altaffiliation can be used to indicate some secondary information
%% such as fellowships. This command produces a non-numeric footnote that is
%% set away from the numeric \affiliation footnotes.  NOTE that if an
%% \altaffiliation command is used it must come BEFORE the \affiliation call,
%% right after the \author command, in order to place the footnotes in
%% the proper location.
%%
%% Use \email to set provide email addresses. Each \email will appear on its
%% own line so you can put multiple email address in one \email call. A new
%% \correspondingauthor command is available in V6.31 to identify the
%% corresponding author of the manuscript. It is the author's responsibility
%% to make sure this name is also in the author list.
%%
%% While authors can be grouped inside the same \author and \affiliation
%% commands it is better to have a single author for each. This allows for
%% one to exploit all the new benefits and should make book-keeping easier.
%%
%% If done correctly the peer review system will be able to
%% automatically put the author and affiliation information from the manuscript
%% and save the corresponding author the trouble of entering it by hand.

\correspondingauthor{Valeriy Vasilyev}

\author[0009-0009-3020-3435]{Valeriy Vasilyev}
\affiliation{ Max-Planck-Institut für Sonnensystemforschung, Justus-von-Liebig-Weg 3, D-37077 Göttingen, Germany \\}
\author[0000-0002-1299-1994]{Timo Reinhold}
\affiliation{ Max-Planck-Institut für Sonnensystemforschung, Justus-von-Liebig-Weg 3, D-37077 Göttingen, Germany \\}
\author[0000-0002-8842-5403]{Alexander I. Shapiro}
\affiliation{ 
Max-Planck-Institut für Sonnensystemforschung, Justus-von-Liebig-Weg 3, D-37077 Göttingen, Germany \\}
\affiliation{ Institute of Physics, University of Graz, 8010 Graz, Austria \\
}
\author[0000-0002-0335-9831]{Theodosios Chatzistergos}
\affiliation{ Max-Planck-Institut für Sonnensystemforschung, Justus-von-Liebig-Weg 3, D-37077 Göttingen, Germany \\}
\author[0000-0002-1377-3067]{Natalie Krivova}
\affiliation{ Max-Planck-Institut für Sonnensystemforschung, Justus-von-Liebig-Weg 3, D-37077 Göttingen, Germany \\}
\author[0000-0002-3418-8449]{Sami K. Solanki}
\affiliation{ Max-Planck-Institut für Sonnensystemforschung, Justus-von-Liebig-Weg 3, D-37077 Göttingen, Germany \\}

\email{
vasilyev@mps.mpg.de
}
%\affiliation{American Astronomical Society \\
%1667 K Street NW, Suite 800 \\

\begin{abstract}
%limited to 150 words!!!!
%The advent of space-based photometric surveys has enabled the measurement of rotation periods for tens of thousands of stars.
The light curves of old G-dwarfs obtained in the visible and near-infrared wavelength ranges are highly irregular. This significantly complicates the detectability of the rotation periods of stars similar to the Sun in large photometric surveys, such as Kepler and TESS. In this study, we show that light curves collected in the ultraviolet wavelength range are much more suitable for measuring rotation periods. Motivated by the observation that the Sun's rotational period is clearly discernible in the UV part of the spectrum, we study the wavelength dependence of the rotational period detectability. 
We employ the  Spectral and Total Solar Irradiance Reconstructions model, SATIRE-S, to characterize the detectability of the solar rotation period across various wavelengths using the autocorrelation technique. 
We find that at wavelengths above 400 nm, the probability of detecting the rotation period of the Sun observed at a random phase of its activity cycle is approximately 20\%. The probability increases to 80\% at wavelengths shorter than 400 nm. These findings underscore the importance of ultraviolet stellar photometry.
\end{abstract}

\keywords{ Solar faculae(1494) ---
Solar cycle(1487) ---
Solar activity(1475) ---
Solar rotation(1524) ---
Stellar rotation(1629) ---
Solar ultraviolet emission (1533) ---
Space telescopes(1547) --- 
Space observatories(1543) ---
Ultraviolet astronomy(1736)	---
Detection(1911) ---
Sunspots(1653) ---
Stellar activity(1580)}


\section{Introduction} \label{sec:intro}
The rotation period is one of the key characteristics of a star. In particular, it defines the efficiency of the stellar dynamo and various manifestations of stellar activity, such as brightness variations, flare frequency, or UV emission \citep{Basri2021}.  Thus, knowledge of the rotation period of a star is important for assessing the conditions on its planets and for distinguishing between stellar and planetary signatures in observations.

The rotation period of a star can be used to derive its  age \citep{Skumanich1972ApJ, Barnes2003ApJ}. This is because stars spin down over their main sequence lifetime due to the loss of angular momentum through magnetized stellar winds. Furthermore, identifying a large sample of stars with near-solar rotation periods is needed for solar-stellar comparison studies and, in particular, for constraining the full range of solar activity and variability \citep{Reinhold2020, Santos2023}.

As the star rotates, active regions on the stellar surface transit across the visible stellar disk, causing quasi-periodic variations in the light curve. These variations allow for detecting rotation periods with the autocorrelation function \citep[ACF, see, e.g.][]{McQuillan2013}, Lomb-Scargle periodogram \citep{Reinhold2013}, or Gaussian process techniques \citep{Angus2018}.
Consequently, the analysis of light curves measured by large space-based telescopes (in particular, by Kepler) has made possible the detection of rotation periods of many tens of thousands of stars \citep{McQuillan2014, Santos2019, Santos2021, Reinhold2013, Reinhold2023}.

However, on slowly rotating stars like the Sun, the lifetime of most of the spots is shorter than the rotation period \citep{DrielGesztelyi2015}. Furthermore, on such stars, the rotational signal from dark spots is partly compensated by the signal from bright faculae, making the detection of rotation periods even more challenging \citep{Shapiro2017, Reinhold2021, Eliana2020, Witzke2020}. 
Consequently, the probability of detecting 
the correct rotation periods of sun-like stars (early G-type stars with sun-like rotation periods) for different inclination angles and metallicities 
%the correct rotation period of a solar twin 
using the commonly applied ACF technique by \cite{McQuillan2013}  is only 3\% \citep{Reinhold2021}. This significantly complicates solar-stellar comparison studies \citep[see e.g.,][]{Reinhold2020,Valera2024}. 

One way to solve this problem is to observe stars at wavelengths where brightness variability is dominated by faculae. Facular features live much longer than spots; for example, solar facular features can persist for multiple solar rotations. As a result, the pattern of the brightness variability caused by faculae is much more regular than that caused by spots. 
For instance, stellar rotation periods could be readily determined from the monitoring of the S-index \citep[see e.g.,][]{Stimets1980, Wright2004, Hempelmann2016, Mittag2017} which is a measure of the emission in the cores of the \caHK lines and is caused by faculae for a wide selection of stars \citep{sowmya2023}.  Similarly, the solar rotation period is prominently reflected in variations of solar UV irradiance, which originates in the facula-dominated upper atmosphere of the Sun. Observations of UV irradiance over time have clearly revealed this periodicity \citep{Rottman1982, London1984, ROTTMAN199937}.  

The facular brightness contrast and, consequently, the facular contribution to brightness variability has a complex wavelength dependence. It generally increases towards the shorter wavelengths,  but it is also strongly amplified by various molecular line systems and atomic lines. Consequently, one can expect a strong dependence of the detectability of the rotation period on the wavelength. Here we study this dependence in the exemplary case of the Sun. We apply the autocorrelation algorithm developed by \cite{McQuillan2014} to time series of solar irradiance  (from ultraviolet to infrared) taken from SATIRE-S reconstruction \citep{SATIRES2014}. The structure of the manuscript is the following. Sect.~\ref{sec:data} describes the solar irradiance time series we use. Sect.~\ref{sec:method} describes the setup of the autocorellation method. In Sect.~\ref{sec:results} we present the results and, finally, we summarize our conclusions in Sect.~\ref{sec:summary}.

\section{Data} \label{sec:data}
We used the solar spectral irradiance (SSI) reconstructed with the Spectral and Total Irradiance REconstruction \citep[SATIRE-S, with ``S'' standing for the Satellite era;][]{Fligge2000, krivova_reconstruction_2003,SATIRES2014} model. 
%SATIRE-S is a physics-based semi-empirical model.
SATIRE-S is a physics-based semi-empirical model, which reproduces the measured variability of the solar total and spectral irradiance on time scales of days to decades \citep{Krivova2006, Unruh2008, Unruh2012, Yeo2015}.
To reconstruct irradiance variations, SATIRE divides the solar surface into four components: quiet Sun, faculae, sunspot umbra, and sunspot penumbra.
The intensity of each component was computed with the radiative transfer code ATLAS9 \citep{kurucz_atlas_1970} from the corresponding semi-empirical model atmospheres \citep{unruh_spectral_1999}.
Continuum observations and magnetograms were used to derive the distribution of sunspots and faculae on the solar surface, respectively.

We used two space-based sets of observations: the Helioseismic and Magnetic Imager \citep[HMI;][]{scherrer_helioseismic_2012} onboard the Solar Dynamics Observatory \citep[SDO;][]{pesnell_solar_2012} and the Michelson Doppler Imager \citep[MDI;][]{scherrer_solar_1995} on board the Solar and Heliospheric Observatory \citep[SoHO;][]{domingo_soho_1995}.
The SoHO/MDI data used here cover the period between 2 February 1999 and 29 April 2010, while after 30 April 2010 the SDO/HMI data were used.
The computed daily SSI covers the spectral range from 115 nm to 160000 nm.
However, here we restrict our analysis to the spectral range 200~nm to 1000~nm, with a typical step of 1--2 nm.

In addition to SSI, we also used individual facular and spot contributions to the SSI variability.
These were computed by considering only either faculae or spots while treating the locations of the other component as the quiet Sun (QS).

\section{Method} \label{sec:method}
\subsection{Time Series Analysis}
To mitigate the impact of long-term variations (solar cycle and longer), we detrend the time series of solar irradiance by subtracting its 81-day moving average. This is done at each wavelength separately.
%%%%%%%%%%%%%%%%%%%%%%%%%%%%%%%%%
\begin{figure*}[ht!]
    \centering
     \includegraphics[width=1.0\textwidth]{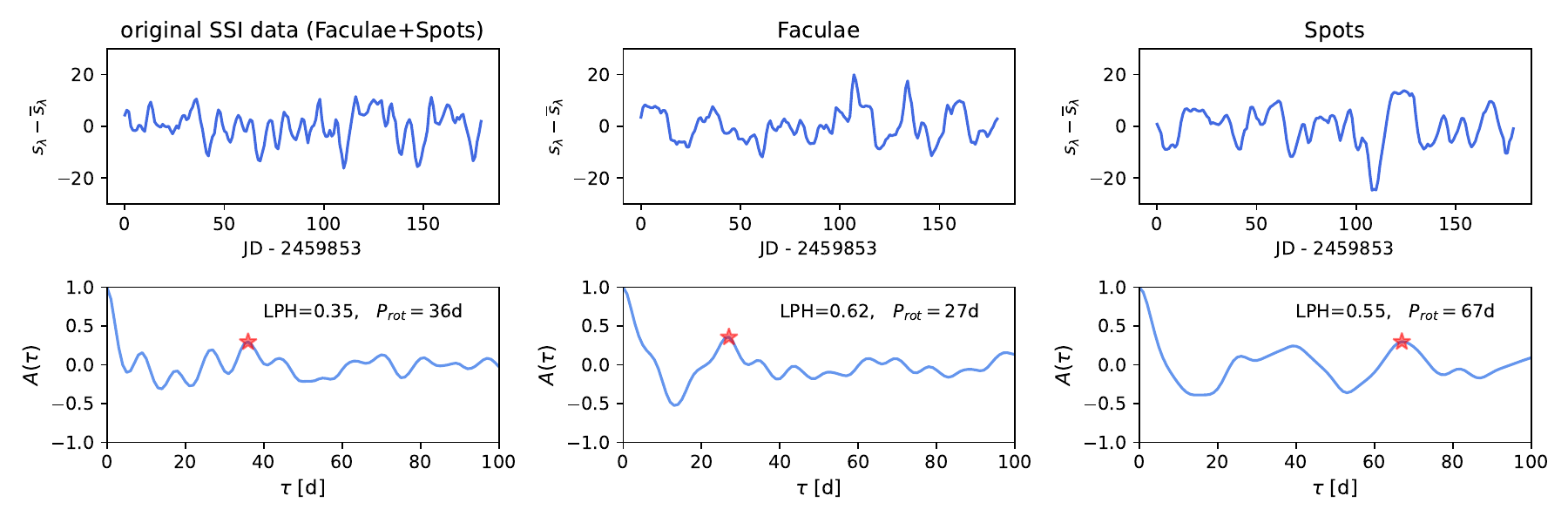}
 \caption{Determination of solar rotation period using the autocorrelation function. \textit{Top panels:} 180-day long time series of SSI averaged over 
 441--442 nm spectral interval (left)  as well as the facular (middle) and spot (right) contributions to this time series. The long-term trends from these three time series were removed by subtracting the 81-day running mean. \textit{Bottom panels:}  Autocorrelation functions computed for the time series shown in the respective top panels. Red markers indicate the highest peaks for each case (at a non-zero time lag).}
    \label{fig:prot-method}
\end{figure*}
%%%%%%%%%%%%%%%%%%%%%%%%%%%%%%%%%%%%%%%%%%%%

From the detrended time series, we select 1000 segments of 180 days each, randomly distributed in time. For each data segment, we compute the ACF.

\subsection{Rotation period determination}
We follow the method by \citet{McQuillan2013, McQuillan2014} used to measure the rotation periods of tens of thousands of Kepler stars. We identify the highest local extrema in the ACF.  However, correlated noise and residual systematics can introduce underlying long-term trends, which means the absolute peak height is no longer a good diagnostic. To overcome this, we focus on the ``local peak height" (LPH), defined as the height of the primary ACF peak with respect to the troughs on each side. To accurately identify real periods and reduce false positives, we establish a threshold of $0.25$ for the LPH. If the highest peak in the ACF exceeds $\mathrm{LPH}>0.25$, we designate the lag of this peak as the rotation period.  We only search for peaks at periods shorter than 70 days \citep{McQuillan2014}. 
Following the definition of \cite{Reinhold2021}, if the highest ACF peak lies between 24 and 30 days, we count it as the true rotation period detection. This interval is centered on the Carrington rotation period ($\approx 27$~days), with the lower boundary of 24 days corresponding roughly to the solar rotation period at the equator and the upper boundary set at 30 days, symmetrically around the mean.%, equidistant from the Carrington period.

Since the time series have a cadence of 1 day, the measured rotation period has an inherent uncertainty of 1 day. Furthermore, time evolution of sunspots and  faculae, introduces additional uncertainties. By allowing a margin of $\pm3$ days around the Carrington period, we account for these uncertainties while maintaining robust period determination.

We applied the method described above to measure the rotation period across each of the 1000 180-day-long detrended SSI segments. To better interpret the results, we also conducted this  analysis on two additional data sets created for the same wavelength range as SSI: one including only a contribution from faculae  and another featuring the spot contribution alone.

In Figure~\ref{fig:prot-method}, we show the rotation period determination in the SSI data and in the data containing only facular and spot contributions, both averaged over the 441–442 nm spectral range and covering the same time interval. Only in the facular data is there a strong peak ($\mathrm{LPH}=0.62$) at 27 days, corresponding to the correct solar rotation period.
%%%%%%%%%%%%%%%%%%%%%%%%%%%%%%%
\section{Results} \label{sec:results}
\begin{figure*}[ht!]
\centering
\includegraphics[width=0.9\textwidth]{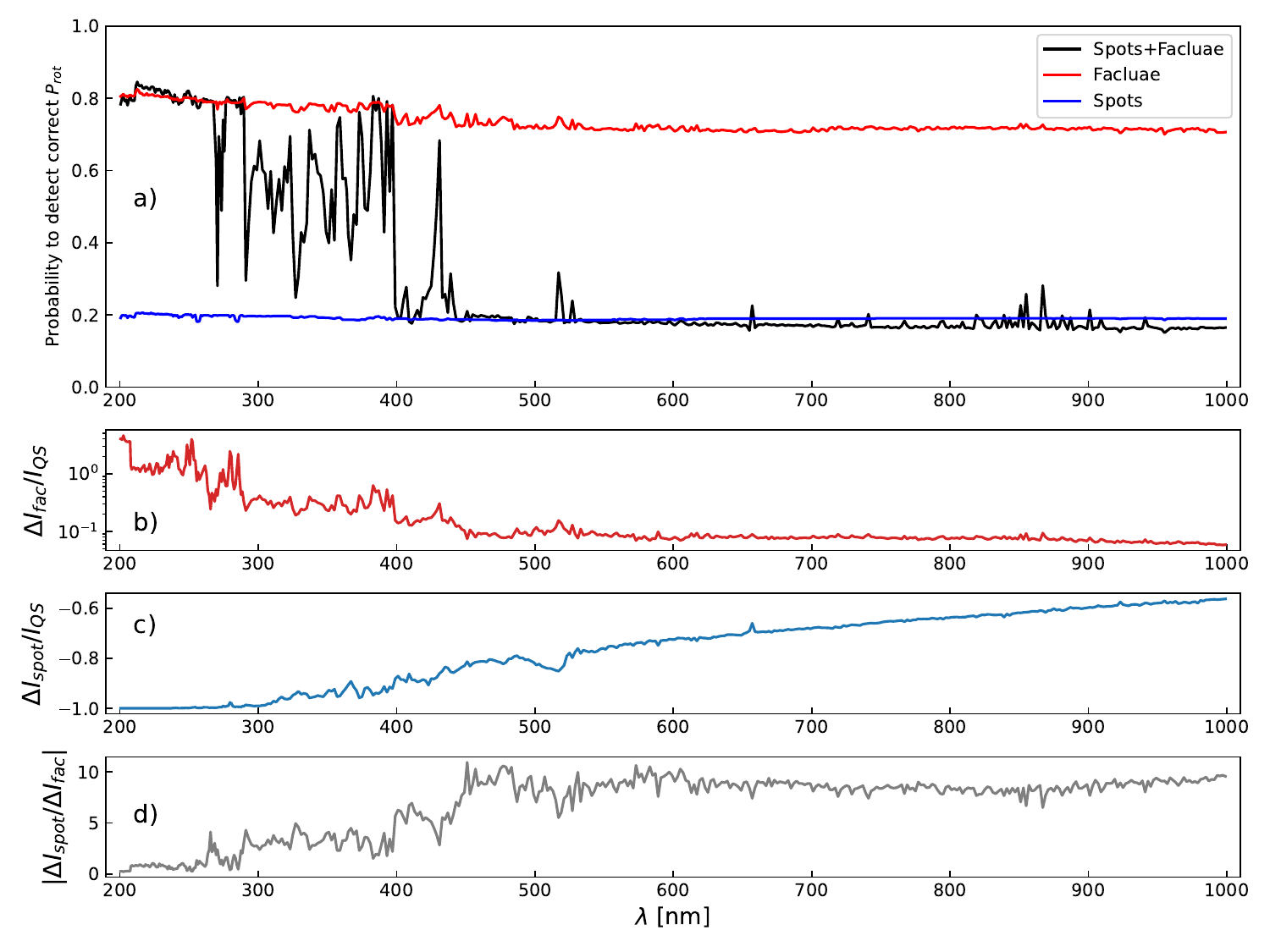}
\caption{Recovery of the solar rotation period at various wavelengths (a). Shown are 
probabilities of detecting the true solar rotation period using the autocovariance technique in the original SSI data (black), its facular (red), and spot (blue) components at a random phase of the activity cycle. For illustrative purposes we also show disc-integrated facular {(b)} and spot {(c)} contrasts, and the spot-to-facular contrast ratio {(d)}.
}
\label{fig:prot-determination-1d}
\end{figure*}

We define the probability of detecting the true rotation period at a given wavelength at a random phase of its activity cycle as the ratio of the number of data segments exhibiting peak heights $\mathrm{LPH}>0.25$ at periods within the 24 to 30 days range to the total number of considered data segments, which is 1000. The probabilities computed for the overall SSI, as well as for the facular and spot contributions to it  are presented in Figure~\ref{fig:prot-determination-1d}a.

In the SSI data, the percentage of detections of the true rotation period depends strongly on the wavelength. For $\lambda<300$~nm, the true period is recovered in approximately $80$\% of cases except for two deep drops at  270 and 290 nm. At wavelengths $\lambda>400$ nm, the probability of detecting the correct rotation period drops to around 20\%. 

To understand these results, we looked at the individual contributions of spots and faculae to the variability. When only the facular component is considered, the probability of recovering the true rotation period is around 80\% for the entire considered wavelength range.  This high probability is due to the lifetimes of faculae, which are longer than the solar rotation period. Modulated with rotation they make light curves more regular, leading to higher LPH values. However, even in such cases, the correct rotation period is not necessarily associated with the highest ACF peak. For example, due to the center-to-limb brightening of faculae, the highest ACF peak can appear at half of the rotation period. 

When only the spot component is considered, the probability is around 20\% and it does not strongly depend on the wavelength, because the spot contrast is less wavelength-dependent than the facular contrast (see Figure~\ref{fig:prot-determination-1d}b and c). Such a low probability (compared to the facular case) is caused by the short lifetime of most spots, which reduces the light curve regularity.

In the SSI data, at wavelength $\approx400$~nm there is a transition from the faculae- to the spot-dominated regime of rotational variability \citep{Shapiro2016} (see also  Figure~\ref{fig:prot-determination-1d}d). 
At wavelengths shorter than $\approx400$~nm the rotational variability is primarily driven by faculae.  The sharp peaks in the 280--400~nm range, correspond to the narrow molecular CN, NH, and OH bands which are within the faculae-dominated regime of solar variability on the rotational timescale due to the strong temperature sensitivity of the molecular lines \citep[see details in ][]{Shapiro2015, Shapiro2016}. The peak at around 430 nm associated with the molecular CH G-band has basically the same cause. 
%%%%%%%%%%%%%%%%%%%%%%%%%%%%%%%%%%%%%%%%%%%%%%%%%%%%%%%%%%%%%
\begin{figure*}[ht!]
    \centering
     \includegraphics[width=\textwidth]{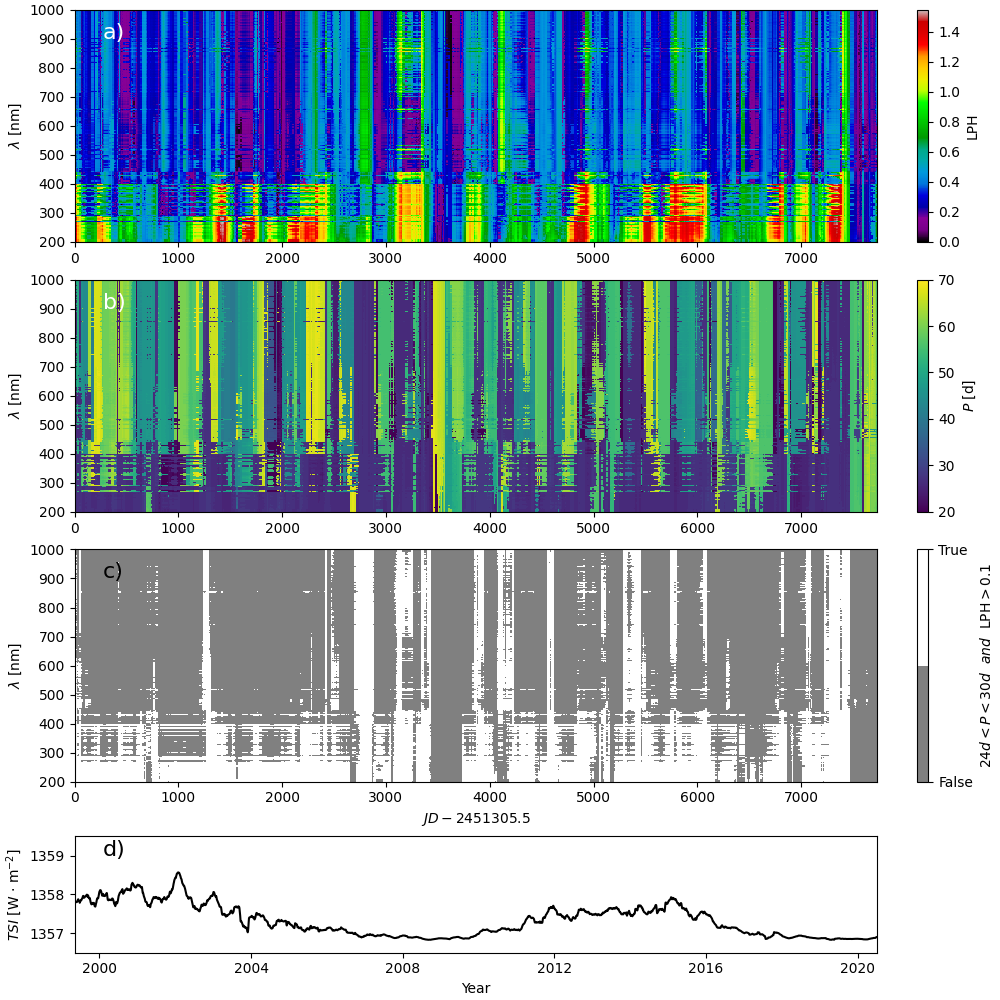}
\caption{Recovery of the solar rotation period as a function of time (abscissa) and wavelength (ordinate). a) Measured local peak heights.
(b) Measured rotation periods.
(c) White patches indicate wavelengths and times where the true rotation period is recoverable, while grey patches represent the wavelengths and times where it is not. d) The TSI time series covering the same time interval for comparison.
%White patches indicate the wavelengths and times where the true rotation period is recoverable, while grey patches represent the wavelengths and times where it is not.  For comparisson, we also show TSI time series in the bottom panel.}
}
\label{fig:prot-determination-2d}
\end{figure*}
%%%%%%%%%%%%%%%%%%%%%%%%%%%%%%%%%%%%%%%%%%%%%%%%%%%%%%%

In Figure~\ref{fig:prot-determination-2d}, we
show the detectability of the solar rotation period, as well as the values of LPHs and the corresponding periods as functions of wavelength and time derived from SSI changes, including contributions from both spots and faculae. At wavelengths $>400$~nm, approximately 60\% of the LPHs exceed the upper threshold, whereby the corresponding periods are randomly distributed across the parameter range. At shorter wavelengths, the LPHs are systematically highest and are mostly close to the true period.  
We found short time intervals when the true solar rotation period can still be measured across all wavelengths (see the $2700-2900$~days and $7200-7500$~days intervals). These short intervals are close to the end of the activity  cycle (see the corresponding TSI time series in the bottom panel of Figure~\ref{fig:prot-determination-2d}).
During these intervals, the facular component is the main source of variability on the rotational timescale at all considered wavelengths. 

During the solar cycle minimum (see the time intervals $3400-3750$~days and $7500-7700$~days),  
there are time intervals, when the correct rotation period cannot be measured at any  wavelength. These intervals correspond to the QS observations, i.e. an absence of spots and a very weak facular component.

All in all, this indicates that detectability of the correct period depends not only on wavelength but also on the spatial coverage of active regions and the faculae to spot ratio that are functions of the activity cycle phase. One straightforward way to significantly increase the chance of recovering the true period is by placing a spectral filter toward wavelengths s shorter than 400 nm, i.e. in the UV (see details in Appendix~\ref{A1})

\section{Discussion and Summary} 
\label{sec:summary}
We studied the detectability of the solar rotation period across various wavelengths using the SATIRE-S model. We found that at the wavelengths observed by the Kepler telescope  (400-900 nm), the correct period can be detected in only $20\%$ of cases (noiseless case).  We note that this value is higher than the 10\% reported in \cite{Reinhold2021} for the noiseless case (see their Table~1). 
This is because we opted to stick to the Sun observed from its equatorial plane. In contrast,  \cite{Reinhold2021} considered the effect of observations from the out of the equatorial plane and also stars with non-solar metallicity. Consequently, the sample of \cite{Reinhold2021} also contained stars with non-solar patterns of variability \citep{Witzke2018, Witzke2020}. 

All in all, the low probability of detecting rotation periods from observations in the visible spectral domain agrees with the explanation given by \cite{Reinhold2020, Reinhold2021}, accounting for why most rotation periods of solar-like stars remain undetected in Kepler and TESS stellar samples (see \cite{vanSaders2019} and \cite{Claytor2024}  for Kepler and TESS estimates, respectively).

In the visible spectral domain, the brightness variability of solar-like stars is dominated by the spot contribution. The main limiting factor for detecting the rotation period is the irregularity of the light curve due to the generally short spot lifetimes. Furthermore, the brightness changes caused by the dark spots and bright faculae partly compensate each other, which further reduces the amplitude of the rotational signal \citep{Shapiro2017, Witzke2020,  Nemec2020}. An exception from this general tendency is epochs of low solar activity when the rotational variability is attributed to facular features. They typically last longer than the solar rotation period, making the light curve pattern more periodic.

We showed that the probability of detecting rotation period strongly increases in the UV spectral domain (namely, shortward of the CH violet system at 430 nm), reaching 80\% at several spectral bands. We believe this finding is of importance for future stellar observations in the UV, such as those by the Ultraviolet Explorer (UVEX) \citep{UVEX}, Mauve \citep{MAUVE},
Large Ultraviolet Optical Infrared Surveyor (LUVOIR) \citep{LUVOIR}, and Wide Field Spectroscopic Telescope (WST) \citep{WST}. 

Another way to increase the detectability of the rotation periods is to improve methods for rotation period detection \citep[see, e.g.][]{Shapiro2020,  ElinanaGPS2020, Santos2021}. For example, \cite{Santos2021} combined wavelet analysis with the ACF method and measured rotation periods for 39592 G- and F-dwarfs and subgiants.  Later, \cite{Reinhold2023} complemented the ACF method with the gradient of the power spectrum of variability \citep[GPS method, see ][]{Shapiro2020} 
to obtain rotation periods of 67163 Kepler stars. In contrast to standard frequency analysis methods, the GPS method  does not rely on the regularity of the light curves and thus is well suited for measuring rotation periods of stars similar to the Sun \citep{ElinanaGPS2020, Reinhold2022ApJ}. 

UV observations have a great potential to detect rotation periods of an even larger sample of solar-like stars. Furthermore, combining periods detected in UV with the GPS  method will allow constraining properties of stellar activity cycles  \citep[e.g., the ratio of the stellar areas covered by facuale and spots]{Eliana2020b, Reinhold2021}.

\newpage
\section{Appendix}
\subsection{Period detectability with rectangular filter} \label{A1}
Here, we consider a rectangular filter with 100\% transmission characterized by two parameters: the central wavelength and the width.  We change the central wavelength of the filter within the range 275--1000~nm with a step of 25~nm and the width within 5--150~nm with a step of 7~nm. We compute the light curve in each filter, apply the time series analysis, measure the rotation period as described above, and compute the probability of detecting the correct rotation period. 

\begin{figure*}[ht!]
\centering
\includegraphics[width=0.9\textwidth]{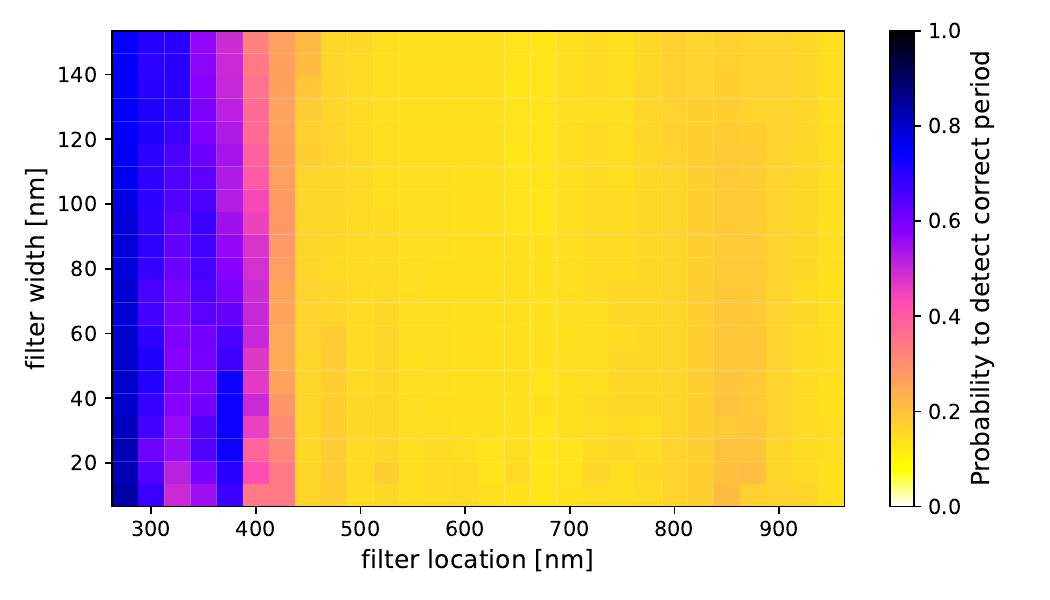}
\caption{Detectability of the solar rotation in photometric observations with a rectangular filter. The probability is shown as a function of the central wavelength and width of the filter.}
\label{fig:prot-determination-2d-filters}
\end{figure*}

In Figure~\ref{fig:prot-determination-2d-filters}, we show the probability of detecting the correct rotation period as a function of filter width and filter location. We find that for all considered filter widths that are centered at wavelength longer than 425~nm the probability  is around 20\%. Centering the filter at wavelengths shorter than 400~nm increases the probability to 40--70\%. Generally, for determining the correct rotation period the filter location plays a bigger role than the filter width.
%%%%%

\begin{acknowledgments}
VV acknowledges support from the Max Planck Society under the grant ``PLATO Science'' and from the German Aerospace Center under  ``PLATO Data Center'' grant   50OO1501.
AIS acknowledges support from the European Research Council (ERC) under the European Union's Horizon 2020 research and innovation program (grant No. 101118581 — project REVEAL).
SKS and TC acknowledge funding from the European Research Council (ERC) under the European Union's Horizon 2020 research and innovation program (grant No. 101097844 — project WINSUN).
\end{acknowledgments}

\software{
matplotlib \citep{matplotlib}, 
numpy \citep{numpy}, 
scipy \citep{scipy}
}



\bibliography{bibiliography}{}
\bibliographystyle{aasjournal}

\end{document}